\documentclass{moriond}
\usepackage{amsmath,amsfonts,latexsym,amssymb}
\usepackage{adjustbox}
\usepackage{color}

\def\be{\begin{equation}}
\def\ee{\end{equation}}
\def\bea{\begin{eqnarray}}
\def\eea{\end{eqnarray}}

\newcommand{\Photo}{}
\newcommand\sss{\scriptscriptstyle}
\newcommand{\OO}{\ensuremath{\mathcal{O}}}
\newcommand{\Op}[1]{\OO_{\sss #1}}
\newcommand{\Opp}[2]{\OO_{\sss #1}^{\sss #2}}
\newcommand{\pdp}{\ensuremath{H^\dagger H}}
\newcommand{\Cp}[1]{C_{\sss #1}}
\newcommand{\Cpp}[2]{C_{\sss #1}^{\sss #2}}
\newcommand{\mw}{\ensuremath{M_{\sss W}}}
\newcommand{\mz}{\ensuremath{M_{\sss Z}}}
\newcommand{\red}[1]{ \textcolor{red}{#1} }

\makeatletter
\newcommand{\oset}[3][0ex]{%
  \mathrel{\mathop{#3}\limits^{
    \vbox to#1{\kern-2\ex@
    \hbox{$\scriptstyle#2$}\vss}}}}
\makeatother

\newcommand{\lra}[1]{\ensuremath{\oset[-0.1ex]{\leftrightarrow}{#1}}}

\begin{document}
\begin{flushright}
KCL-PH-TH/2021-32
\end{flushright}

\vspace*{4cm}
\title{Top quark electroweak interactions}

\author{ K. Mimasu }

\address{Theoretical Particle Physics and Cosmology Group, Department of Physics, King’s College London, London WC2R 2LS, UK}

\maketitle\abstracts{
I discuss how modified EW interactions lead to energy growth in $2\to 2$ scattering amplitudes and how the framework of the Standard Model Effective Field Theory can be used in conjunction with the Goldstone boson equivalence theorem to better understand the origin of such unitarity violating behaviour. I quantify how applications in constraining top quark electroweak couplings are especially motivated due to our relatively poor understanding of them. I present some highlights of a systematic study of energy growth in EW top scattering amplitudes with a focus on how such effects can be searched for in realistic collider processes, demonstrating some promising avenues for future study.}

\section{Introduction}\label{sec:intro}

One of the great challenges facing the current generation of high energy physicists is to uncover the origin of electroweak (EW) symmetry breaking (EWSB). Alongside direct searches for new states at the TeV scale, careful study of the properties of the relevant Standard Model (SM) degrees of freedom is essential. The particles that are most strongly implicated in this mechanism are those with the greatest mass, coupling most strongly with the Higgs field. This naturally includes the Higgs boson itself as well as the $W$ and $Z$ bosons and the top quark. By precisely characterising the interactions among this sector, we may find indirect evidence of new physics that has thus far eluded direct searches, but nevertheless left an imprint in couplings at lower energies. 
%
This imprint can perturb the delicate balance that conserves unitarity and renormalisability in the SM. 
Indeed, the EW sector as we know it is the unique, weakly-coupled realisation of our observed low energy field content, that respects unitarity up to arbitrarily high energies~\cite{Cornwall:1973tb,LlewellynSmith:1973yud,Cornwall:1974km}. Spoiling any of the tight relations between couplings and/or masses at low energies, leads to unacceptable energy growth in scattering amplitudes. This necessarily implies that the theory now has a finite range of validity and that new, heavy states should exist.

The most well known application of these ideas considered the scattering of longitudinally polarised of $W$-bosons, $W_{\sss L}W_{\sss L}\to W_{\sss L}W_{\sss L}$~\cite{LlewellynSmith:1973yud,Lee:1977eg,Lee:1977yc}, and allowed us to bound the scale of EWSB from above~\cite{Lee:1977yc,Lee:1977eg}, guaranteeing the discovery of new physics (which turned out to a very Higgs-like scalar) at the LHC. 
Many other amplitudes display similar cancellations in the SM as a consequence of gauge symmetry and linearly realised EWSB. These can be viewed as testing grounds to search for signs of new physics via unitarity-violating behvaiour. The marriage of precision and high energy measurements seeking energy growth in scattering amplitudes lends itself to an Effective Field Theory (EFT) description. Assuming linearly realised EWSB points to the Standard Model EFT (SMEFT) as the ideal candidate theory in which to interpret LHC data to search for indirect hints of heavy new physics at the TeV scale. In the following, I will present highlights of a broad study of unitarity violation in EW top scattering amplitudes and the associated collider phenomenology~\cite{Maltoni:2019aot}.

\section{The top window}\label{sec:topwindow}
Many interesting, fermionic scattering amplitudes with potential unitarity-violation turn out to be proportional to the fermion mass itself. The top quark is therefore a particulary interesting candidate to study. From a theoretical point of view, the top is often considered to be a promising window into physics beyond the SM in connection to the hierarchy and flavor puzzles. Experimentally, the LHC copiously produces top quarks and is beginning to pin down even its elusive, EW couplings. That said, these interactions, particulary the neutral couplings, are still relatively poorly known. This can be quantified in the SMEFT framework, in which higher-dimension operators parametrise the modification of SM interactions. The leading, operators affecting top EW scatterings appear at dimension-6 and are given in Tab.~\ref{tab:ops}, employing the so called Warsaw basis~\cite{Buchmuller:1985jz,Grzadkowski:2010es}. A 
 $U(3)_{\ell}\times U(3)_{e}\times U(3)_{d}\times U(2)_{q}\times U(2)_{u}$ flavour symmetry has been assumed, to single out modified top quark couplings~\cite{AguilarSaavedra:2018nen}.
\begin{table}[h!]
\renewcommand{\arraystretch}{1.4}
\centering
\caption{\label{tab:ops}
SMEFT operators describing new interactions involving the EW and top quark 
sectors, consistent with a $U(3)^3\times U(2)^2$ flavour symmetry.  
$Q,\,t$ and $b$ denote the third generation components of $q,\,u$ and $d$.}
\begin{tabular}{|ll|ll|}
    \hline
     $\Op{W}$&
     $\varepsilon_{\sss IJK}\,W^{\sss I}_{\mu\nu}\,
                             {W^{{\sss J},}}^{\nu\rho}\,
                             {W^{{\sss K},}}^{\mu}_{\rho}$&
     $\Op{tH}$&
     $\left(\pdp\right)\bar{Q}\,t\,\tilde{H} + \text{h.c.}$
     \tabularnewline
     $\Op{H W}$&
     $\left(\pdp\right)W^{\mu\nu}_{\sss I}\,
                                    W_{\mu\nu}^{\sss I}$&
     $\Op{tW}$&
     $i\big(\bar{Q}\sigma^{\mu\nu}\,\tau_{\sss I}\,t\big)\,
     \tilde{H}\,W^I_{\mu\nu}
     + \text{h.c.}$
     \tabularnewline
     $\Op{H B}$&
     $\left(\pdp\right)B^{\mu\nu}\,
                                    B_{\mu\nu}$&
     $\Op{tB}$&
     $i\big(\bar{Q}\sigma^{\mu\nu}\,t\big)
     \,\tilde{H}\,B_{\mu\nu}
     + \text{h.c.}$
     \tabularnewline
     \cline{3-4}
     $\Op{H WB}$&
     $(H^\dagger \tau_{\sss I}H)\,B^{\mu\nu}W_{\mu\nu}^{\sss I}\,$&
     $\Op{H Q}^{\sss(3)}$&
     $i\big(H^\dagger\!\lra{D}_\mu \!\tau_{\sss I}H\big)
     \big(\bar{Q}\,\gamma^\mu\,\tau^{\sss I}Q\big)$
     \tabularnewline

     $\Op{H D}$&
     $(H^\dagger D^\mu H)^\dagger(H^\dagger D_\mu H)$&
     $\Op{H Q}^{\sss(1)}$&
     $i\big(H^\dagger\!\lra{D}_\mu \!H\big)
     \big(\bar{Q}\,\gamma^\mu\,Q\big)$
     \tabularnewline
     $\Op{H \square}$&
     $(\varphi^\dagger\varphi)\square(\varphi^\dagger\varphi)$ &
     $\Op{H t}$&
     $i\big(H^\dagger\!\lra{D}_\mu\! H\big)
     \big(\bar{t}\,\gamma^\mu\,t\big)$
      \tabularnewline
      &&     
      $\Op{H tb}$&
     $i\big(\widetilde{H}^\dagger {D}_\mu\,H\big)
     \big(\bar{t}\,\gamma^\mu\,b\big)
     + \text{h.c.}$
     \tabularnewline
      \hline
  \end{tabular}
\end{table}
The right-hand column of operators leads to modifications of SM top-quark interactions with all EW gauge bosons apart from the photon, whose interactions are protected by $U(1)_{\sss EM}$ being a good symmetry of the low energy theory. Additionally, weak dipole interactions are also included for the top quark which do induce a modified $t\bar{t}\gamma$ vertex. The left hand column lists purely bosonic operators, that modify triple and quartic gauge couplings as well as gauge-Higgs interactions.

\begin{center}
\begin{figure}[h!]
  \adjincludegraphics[{width=0.9\linewidth,left}]{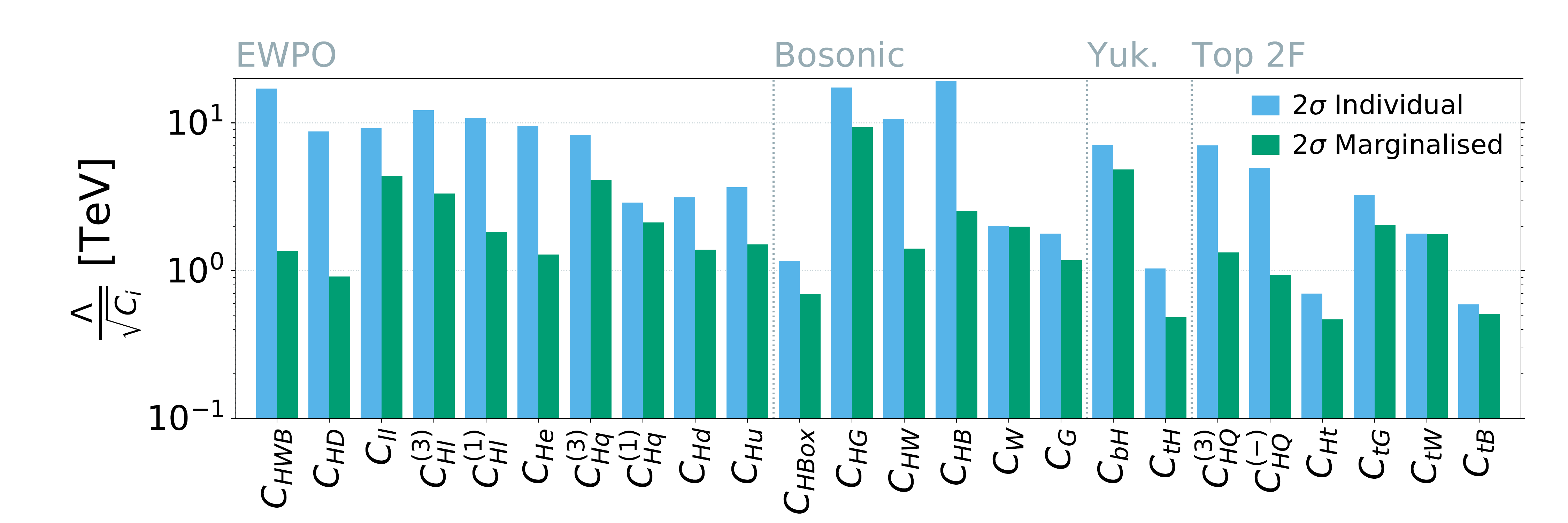}
  \caption{\label{fig:fitresult}
Results of a recent, global SMEFT analysis of Higgs, EW precision, diboson and top 
quark data collected at LEP, Tevatron and the LHC. The
bars indicate the $2\sigma$ sensitivity obtained on each dimensionful operator
coefficient, $C_i/\Lambda^2$, either individually (blue), assuing all others
are zero, or marginalising (green) over the other degrees of freedom. The
values are converted into a scale that corresponds to the value of $\Lambda$
assuming $C_i=1$.
  }
\end{figure}
\end{center}

The level of model-independent sensitivity that we have to SMEFT deformations must be quantified in a holistic way, combining data from a maximum of available experiments. The results of a recent global fit~\cite{Ellis:2020unq} are summarised in Fig.~\ref{fig:fitresult}, which indicates the limits on a subset of the 34 degrees of freedom included in terms of the new physics scale accessed, assuming the Wilson coefficients, $C_i$, are set to 1.
One can clearly see that the top quark Yukawa, $\Cp{tH}$, and neutral EW couplings, $\Cp{Ht}$ (right-handed top-Z), $\Cpp{HQ}{(-)}\equiv \Cpp{HQ}{(1)}-\Cpp{HQ}{(3)}$ (left-handed top-Z) and $\Cp{tB}$ (neutral weak dipole) are among the worst constrained, both individually and globally. On one hand, this suggests that there is still significant room for new physics to hiding in the top sector. On the other, the fact that the global analysis already includes measurements of top pair production in association with a Higgs boson ($t\bar{t}H$), as well as top pair and single top production in association with a $Z$ boson ($t\bar{t}Z$ and $t{Z}j$) underlines the need to look beyond these canonical channels to pin down this sector of couplings. This is especially important given that many of these measurements are already becoming systematics dominated.

\section{High energy EW top scattering}\label{sec:HEEWTS}
One of the main limitations behind the current measurements that are being used to constrain neutral top quark couplings is the lack of energy growth in the underlying amplitudes. For example, although $t\bar{t}H$ is sensitive to the Yukawa operator, $\Cp{tH}$, since it only shifts the top-Higgs coupling, only the inclusive rate is modified, with no interesting kinematical deviations from the SM. No amount of differential measurements will improve the constraints in this channel and systematic uncertainties will fundamentally limit the sensitivity. The same is true for the effect of $\Cpp{HQ}{(-)}$ and $\Cp{Ht}$ on $t\bar{t}Z$. 

This is in conflict with the expectation that higher dimensional operators  yield energy-growing effects that should be exploited at high energy colliders to maximise sensitivity. On dimensional grounds, the general contribution of dimension-6 operator coefficients $C_{x}$ to, \emph{e.g}, a $2\to 2$ scattering amplitude takes the following form:
\begin{align}
    \mathcal{A}_{2\to 2}=\mathcal{A}_{SM}\left(
    1 + C_i\frac{v^2}{\Lambda^2} + C_j\frac{vE}{\Lambda^2} 
      + C_k\frac{E^2}{\Lambda^2}
    \right).
\end{align}
Where the SM amplitude, $\mathcal{A}_{SM}$, has bee factored out to indicate their relative impact. The energy-growing behaviour can range from non-existent to quadratic, with the $t\bar{t}H/Z$ examples belonging to the former case. These processes can, in some sense, be viewed as sub-optimal probes of the couplings at hand, since there must always be \emph{some} amplitude that displays the maximal growth relative to the SM. Contrary to the energy-independent effects, energy-growing modifications will never be limited by systematic uncertainties, since one can always envisage measuring more and more differentially at higher energies, as more statistics are collected. This is reflected in the oft-cited mantra that ``energy helps accuracy''~\cite{Farina:2016rws} for EFT searches at the LHC; it emphasises the point that high energy probes of energy-growing effects can reach comparable sensitivity to low-scale precision data, even with significantly less precise measurements, since their relative sizes are larger at these scales.

The aforementioned unitarity-violating amplitudes correspond exactly to the new physics effects we are after. Consider the example of $bW\to tZ$ scattering, assuming generic left/right-handed top-Z couplings $g_{L/R}$ and $WWZ$ interaction ${g_{WZ}}$. Taking the high energy limit in the Mandelstam variables, $s,t\gg v$, several helicity amplitudes involving longitudinal gauge bosons have energy-growing components, including the maximally-growing:
\begin{align}
    \mathcal(-,0,-,0)\sim\sqrt{s(s+t)}(g_L+g_{WZ}).
\end{align}
The arguments of the amplitude indicate the helicities and polarisations of the $b,W,t$ and $Z$, respectively. The simultaneous appearance of gauge-fermion and gauge self-interactions highlights the interconnectedness predicted by the underlying non-Abelian gauge theory. In this, and all similar amplitudes, the energy-growth cancels when the couplings are assigned their  SM value.

The manifest gauge invariant construction of the SMEFT provides a more intuitive understanding for these effects, when combined with the powerful Goldstone boson equivalence theorem~\cite{Cornwall:1974km}, which identifies amplitudes involving longitudinal gauge bosons, in the high energy limit, with those involving the corresponding Goldstone degree of freedom from the Higgs doublet. Higher dimension operators that modify the (dimension-4) SM couplings must always come with additional powers of Higgs fields taking on their vacuum-expecatation-value, $v$, in order to lower the operator dimension. Away from unitary gauge, this corresponds to new interactions involving Goldstone modes~\cite{Henning:2018kys}. Taking the Higgs current component of $\Op{Ht}$ as an example, we see that in addition to the $Z$ boson field, it contains many other terms 
\begin{align}
    i\big(H^\dagger\!\!\lra{D}_\mu\!\!H\big)\supset v^2Z^\mu + G^{\sss 0}\!\lra{\partial}_\mu h + 2\mz h Z_\mu +\big(i G^-\partial_\mu G^+ + 2i\mw G^- W^+_\mu  + \text{ h.c.}).
\end{align}
These generate contact terms with the right handed top quark current that can be exactly mapped to energy-growth in top EW scattering amplitudes. Analogous arguments can be extented to the complete set of such scatterings to systematically identify promising channels to improve our sensitivity to the operators responsible for EW top quark couplings~\cite{Farina:2012xp,Dror:2015nkp,Maltoni:2019aot}.

\begin{table}
{\footnotesize
\setlength{\tabcolsep}{4pt}
\renewcommand{\arraystretch}{1.2}
\begin{center}
\caption{\label{tab:helamp_summary}
Maximal energy growths induced by each operator on the set of top scattering amplitudes considered. `$-$' denotes either no contribution or no energy growth and the red entries denote the fact that the interference between the SMEFT and the SM amplitudes also grows with energy.
}
  \begin{tabular}{|c|c|c|c|c|c|c|c|c|c|c|c|c|c|}
 \hline
                          & $\Op{H D}$ & $\Op{H \Box}$ &  $\Op{H B}$ & $\Op{H W}$  & $\Op{H WB}$  & $\Op{W}$ & $\Op{t H}$ & $\Op{tB}$ & $\Op{tW}$ & $\Op{H Q}^{\sss (1)}$ & $\Op{H Q}^{\sss (3)}$ &  $\Op{H t}$ & $\Op{H tb}$ 
 \tabularnewline\hline
 
 $b\,W\to t\,Z$           & $E$           & $-$              & $-$            & $-$            & $E$             & $E^2$    & $-$           & $E^2$     & $E^2$     & $E$                      & \red{$E^2$}              & $E$            & $E^2$           
 \tabularnewline\hline

 $b\,W\to t\,\gamma$      & $-$           & $-$              & $-$            & $-$            & $E$             & $E^2$    & $-$           & $E^2$     & $E^2$     & $-$                      & $-$                      & $-$            & $-$           
 \tabularnewline\hline
 
 $b\,W\to t\,h$           & $-$           & $-$              & $-$            & $E$            & $-$             & $-$      & $E$           & $-$       & $E^2$     & $-$                      & \red{$E^2$}              & $-$            & $E^2$            
 \tabularnewline\hline

 $t\,W\to t\,W$           & $E$           & $E$              & $-$            & $E$            & $E$             & $E^2$    & $E$           & $E$       & $E^2$     & \red{$E^2$}              & \red{$E^2$}              & \red{$E^2$}   & $-$            
 \tabularnewline\hline
 
 $t\,Z\to t\,Z$           & $E$           & $E$              & $E$            & $E$            & $E$             & $-$      & $E$           & $E^2$     & $E^2$     & $E$                      & $E$                      & $E$      & $-$                   
 \tabularnewline\hline
 
 $t\,Z\to t\,\gamma$      & $-$           & $-$              & $E$            & $E$            & $E$             & $-$      & $-$           & $E^2$     & $E^2$     & $-$                      & $-$                      & $-$    & $-$                    
 \tabularnewline\hline
 
 $t\,\gamma\to t\,\gamma$ & $-$           & $-$              & $E$            & $E$            & $E$             & $-$      & $-$           & $E$       & $E$       & $-$                      & $-$                      & $-$      & $-$                  
 \tabularnewline\hline

  $t\,Z\to t\,h$           & $E$           & $-$              & $E$            & $E$            & $E$             & $-$      & $E$           & $E^2$     & $E^2$     & \red{$E^2$}              & \red{$E^2$}              & \red{$E^2$}    & $-$           
  \tabularnewline\hline
  
  $t\,\gamma\to t\,h$      & $-$           & $-$              & $E$            & $E$            & $E$             & $-$      & $-$           & $E^2$     & $E^2$     & $-$                      & $-$                      & $-$            & $-$
  \tabularnewline\hline

 $t\,h\to t\,h$            & $E$           & $E$              & $-$            & $-$            & $-$             & $-$      & $E$           & $-$       & $-$       & $-$                      & $-$                      & $-$            & $-$           
  \tabularnewline\hline
  \end{tabular}
\end{center}
\renewcommand{\arraystretch}{1.}
}
\end{table}

Tab.~\ref{tab:helamp_summary}, summarises the key findings of our systematic study of energy growth in SMEFT contributions to ten  $2\to 2$ scattering amplitudes involving at least one top quark and a mixture of gauge/Higgs bosons. It shows the maximum energy growths induced by our set of SMEFT operators, synthesised from a complete calculation of all helicity amplitudes. Energy-growing interference with the SM is especially interesting, since it  corresponds to the leading new physics contribution in the EFT expansion. However, such behaviour is actually quite rare, being only present for current operators in the fully longitudinal gauge boson configurations. The dipole operators, for example, consistently predict maximal growth in the transverse-longitudinal configurations, whose SM amplitudes decrease with energy, leading to relatively suppressed interference. These findings are consistent with the recent demonstration of non-interference at high energies due to helicity selection rules~\cite{Azatov:2016sqh}. 

\section{Embedding the amplitudes into collider processes}\label{subsec:pheno}
Equipped with this information, our task is to determine how best to access such scattering amplitudes in high energy collider experients. Since neither the top, nor the EW gauge bosons can be prepared as an initial state at the LHC, we are forced to seek higher multiplicity EW processes with final states comprising one or two top quarks and EW gauge and/or Higgs bosons and possibly forward jets. 
\begin{figure}
    \centering
\begin{tabular}{ccc}
  \includegraphics[width=0.25\linewidth]{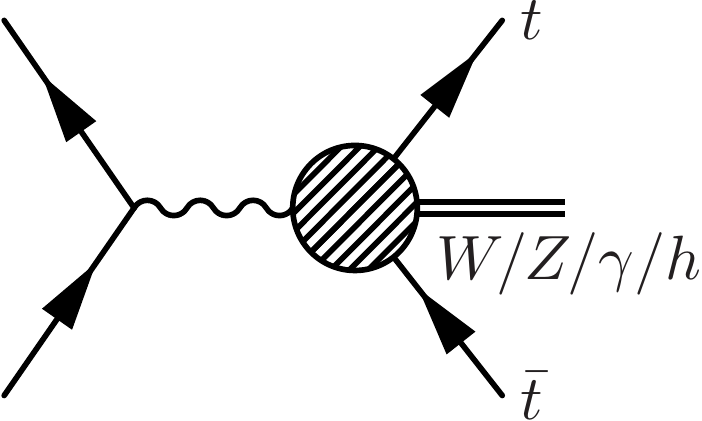} &
  \includegraphics[width=0.25\linewidth]{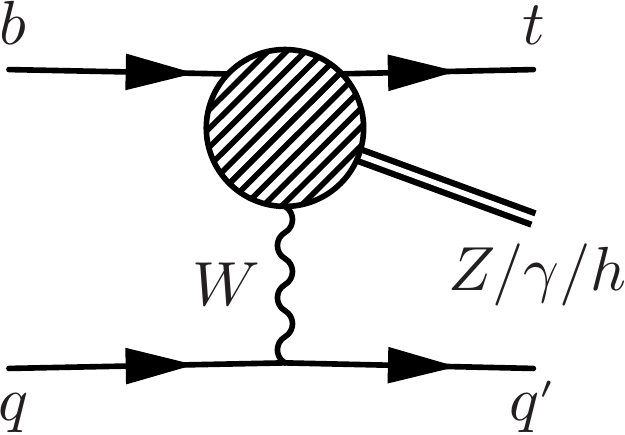} &
  \includegraphics[width=0.25\linewidth]{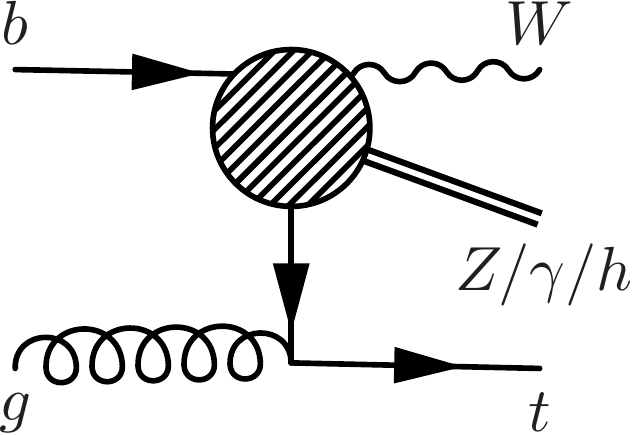} \tabularnewline
  a) $t\bar{t}X$ & b)  $tXj$ & c) $tWX$ \tabularnewline
  \includegraphics[width=0.25\linewidth]{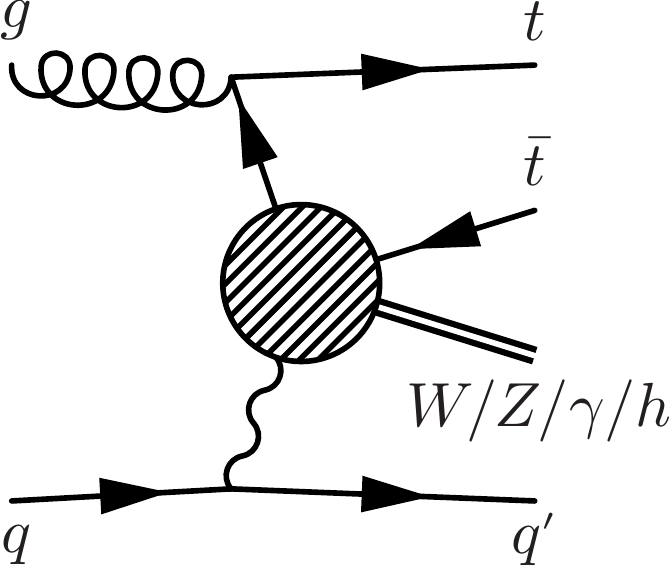} &
  \includegraphics[width=0.25\linewidth]{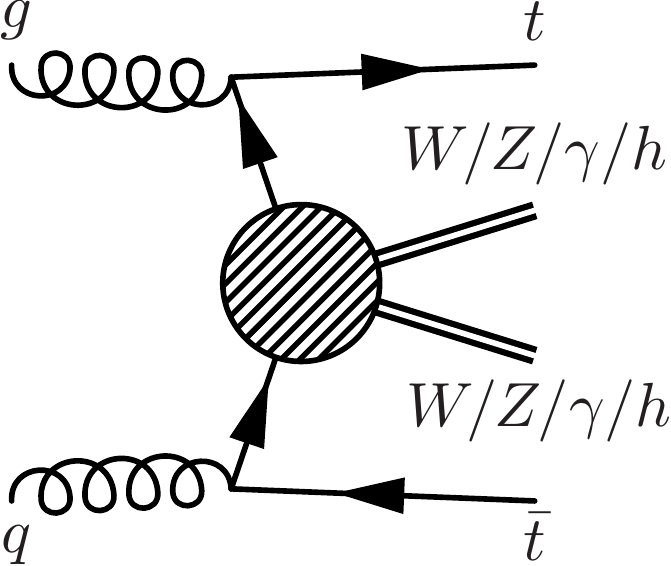} &
  \includegraphics[width=0.23\linewidth]{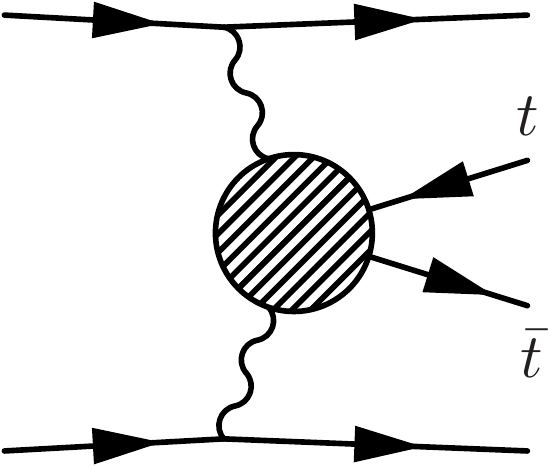} \tabularnewline
  d) $t\bar{t}Xj$ & e)  $ttXY$ & f) VBF \tabularnewline
\end{tabular}
\caption{\label{fig:topologies}
Generic topologies of hadron collider processes than embed top EW scattering amplitudes. 
}
\end{figure}
Fig.~\ref{fig:topologies} depicts an array of hadron collider topologies that can embed the amplitudes of interest, and were investigated in our work. Using the \verb|SMEFTatNLO|~\cite{Degrande:2020evl} model in \verb|Madgraph5_aMC@NLO|~\cite{Alwall:2014hca}, we studied the impact of SMEFT operators at leading order, with respect to the SM predictions, on the cross sections of an array of relevant processes, separating the interference and pure EFT contributions. This part of the study only considered the top quark operators in the right column of Tab.~\ref{tab:ops}, assuming the bosonic operators would be better constrained elsewhere. In order to identify the presence (or not) of energy growth, we computed the relative impact on the inclusive cross section, and after imposing a strong high energy cut, typically on the transverse momentum of the two outgoing particles that participate in the underlying $2\to 2$ scattering. Here, I will restrict the discussion to the example of single top production in association with a $W$ and $Z$ boson, referring the interested reader to the original publication for the full set of processes and phenomenological considerations. 

\begin{figure}[h!]
\begin{center}
\includegraphics[width=0.75\linewidth]{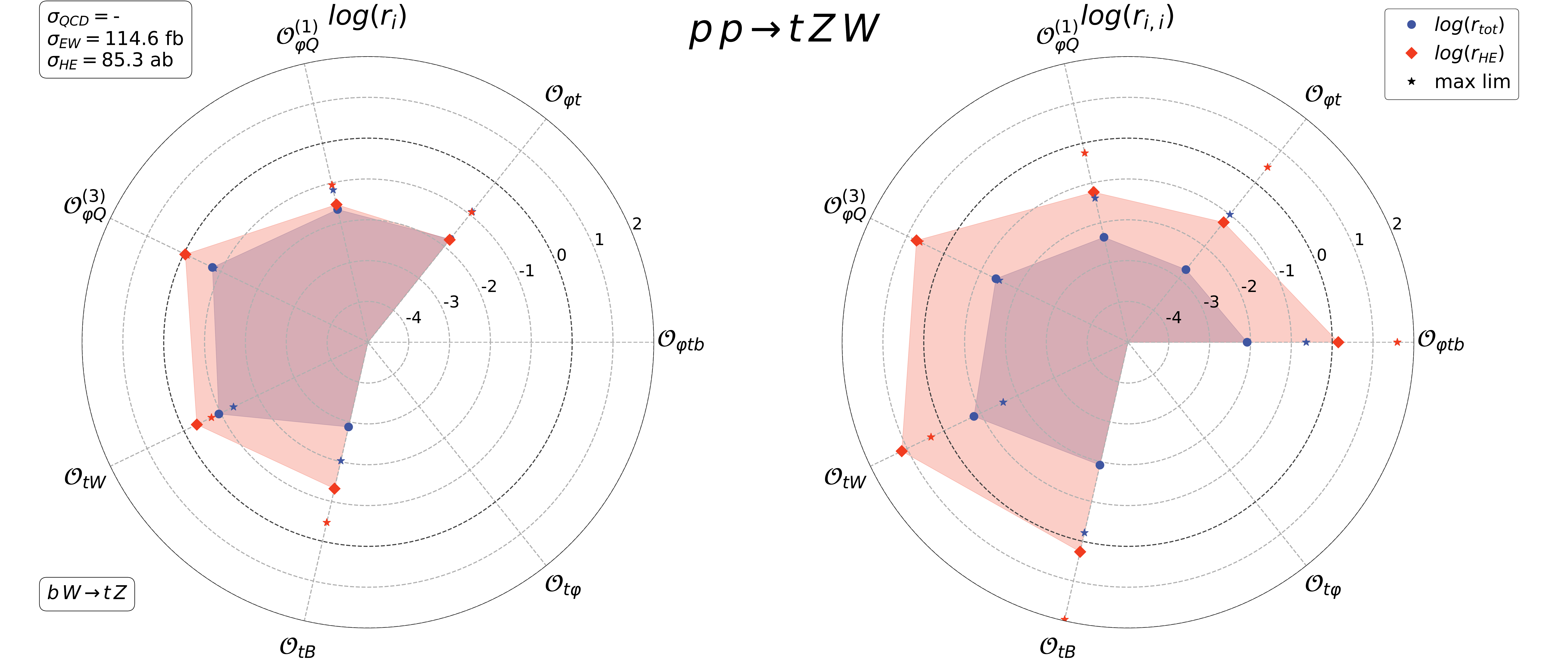}
\caption{\label{fig:radar}
Radar plot for the $p \, p \to t \, W \, Z$ process at the 13 TeV LHC.
The left and right figures show the impact of each operator, for a Wilson coefficient of 1 TeV$^{-2}$, at linear and quadratic level, relative to the SM EW contribution, $\sigma_{\sss EW}$. Dots indicate the impact at inclusive level (blue) and in a high energy region of phase space (red), while stars denote the corresponding prediction when saturating recent individual limits on the coefficients.
}
\end{center}
\end{figure}

$tWZ$ proceeds at the LHC via the topology shown in Fig.~\ref{fig:topologies}$c)$. It embeds the $b\,W\to t\,Z$ subprocess, which is sensitive to 6 of the 7 top quark operators, with the exception of $\Op{tH}$. The inclusive rate in the SM is around 115 fb, which shows an similar suppression with respect to $tZj$ (620 fb) to that of $W$-associated with respect to $t$-channel single top production. The experimental challenge for this channel is its similarity to $t\bar{t}X$, which is analogous to the overlap between the $tW$ and $t\bar{t}$ final states. Nevertheless, the SMEFT-dependence shown in Fig.~\ref{fig:radar}, displays a number of benefits. 
Firstly, by looking at the impact of the high-energy cut ($p_T^{W,Z}$ $>$ 500 GeV), one recognises all of the expected energy-growing behaviour present in the sub-amplitude calculations of Tab.~\ref{tab:helamp_summary}. Most importantly the interference contribution of $\Opp{HQ}{(3)}$ is found to grow with energy, unlike in the case of the $tZj$ process, in which a number of unexpected cancellations, including this one, have been found~\cite{Degrande:2018fog}. Furthermore, the relative impact of the high energy cuts suggest a more pronounced energy growth than for other processes probing $bW\to tZ$ scattering. Second, having both gauge bosons in the final state opens the possibility of measuring both the polarisations and therefore singling out the interesting, longitudinal configurations to enhance EFT sensitivity. Finally, the in contrast to the $tZj$ process, $tWZ$ does not receive contributions from four-fermion operators that could potentially dilute the sensitivity to top EW couplings. The $tWZ$ channel has revealed itself as a very interesting candidate for probing EW top couplings through unitarity violation in high energy/multiplicity scattering, and certainly warrants more detailed study and to eventually be measured during LHC Run 3.

\section{Summary and conclusions}\label{subsec:pheno}
\begin{figure}[h!]
\begin{center}
\includegraphics[width=0.8\linewidth]{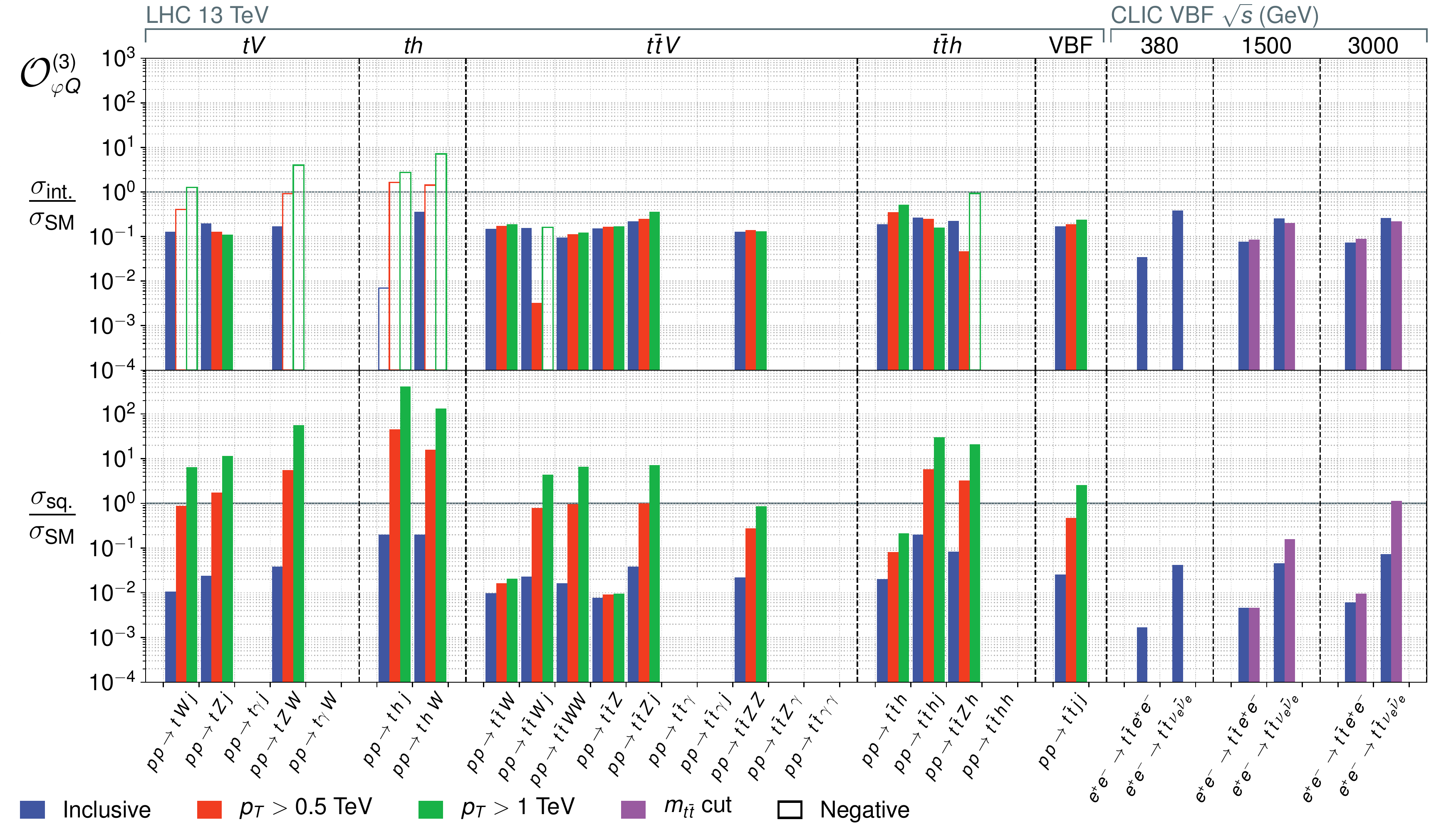}
\caption{\label{fig:moneyplot}
Summary of relative impact on collider processes for $\Opp{\phi Q}{(3)}$ assuming a Wilson coefficient of 1 TeV$^{-2}$. The upper row shows the interference contribution while the lower row, the quadratic piece. The multiple data points per process denote, from left to right, starting with the inclusive rate, the impact as a function of cuts that access increasingly higher energies. Filled and unfilled bars denote constructive and destructive interference terms respectively}
\end{center}
\end{figure}

Top EW couplings are one of the crucial exploration frontiers for LHC physics. I have discussed how, while they may be one of the best places that indirect effects of new physics could hide, they remain relatively poorly known. The SMEFT framework not only allows us to quantify this lack of knowledge in a model independent way but also allows us to understand how modifications to these couplings can lead to unitarity violating behaviour in scattering amplitudes, analogous to those that were extensively studied to pin down the scale of EWSB. These proceedings detail a few highlights of a systematic study of top EW scattering in the SMEFT and potential phenomenological applications at colliders, which identified a number of interesting channels, including $tWZ$ production at the LHC. To highlight the enticing possibilities of such a programme, Fig.~\ref{fig:moneyplot} summarises the high-energy sensitivities to the $\Opp{HQ}{(3)}$ operator, obtained for the 28 collider processes considered in our study, covering high-energy proton-proton and electron-positron colliders. High energy and multiplicity top quark processes are therefore a promising avenue to exploit energy-growing effects to maximise our sensitivity to heavy new physics, and go beyond traditional ways to measure top EW couplings.

\section*{Acknowledgments}
I would like to thank the organisers of the Rencontres de Moriond for the invitation to share my work and for delivering a first and very successful virtual edition.
This work was supported by a Marie Sk{\l}odowska Curie individual Fellowship of the European Commission's Horison 2020 research and innivation programme under contract number 707983, the F.R.S.-FNRS with the EOS - be.h project n. 30820817 and MISU convention F.6001.19, and by the UK STFC via grant ST/T000759/1.


\section*{References}

\begin{thebibliography}{99}
\bibitem{Cornwall:1973tb}
J.~M.~Cornwall, D.~N.~Levin and G.~Tiktopoulos,
Phys. Rev. Lett. \textbf{30} (1973), 1268-1270

\bibitem{LlewellynSmith:1973yud}
C.~H.~Llewellyn Smith,
Phys. Lett. B \textbf{46} (1973), 233-236

\bibitem{Cornwall:1974km}
J.~M.~Cornwall, D.~N.~Levin and G.~Tiktopoulos,
Phys. Rev. D \textbf{10} (1974), 1145
[erratum: Phys. Rev. D \textbf{11} (1975), 972]

\bibitem{Lee:1977eg}
B.~W.~Lee, C.~Quigg and H.~B.~Thacker,
Phys. Rev. D \textbf{16} (1977), 1519

\bibitem{Lee:1977yc}
B.~W.~Lee, C.~Quigg and H.~B.~Thacker,
Phys. Rev. Lett. \textbf{38} (1977), 883-885

\bibitem{Maltoni:2019aot}
F.~Maltoni, L.~Mantani and K.~Mimasu,
JHEP \textbf{10} (2019), 004
[arXiv:1904.05637 [hep-ph]].

\bibitem{Buchmuller:1985jz}
W.~Buchmuller and D.~Wyler,
Nucl. Phys. B \textbf{268} (1986), 621-653

\bibitem{Grzadkowski:2010es}
B.~Grzadkowski, M.~Iskrzynski, M.~Misiak and J.~Rosiek,
JHEP \textbf{10} (2010), 085
[arXiv:1008.4884 [hep-ph]].

\bibitem{AguilarSaavedra:2018nen}
J.~A.~Aguilar-Saavedra, C.~Degrande, G.~Durieux, F.~Maltoni, E.~Vryonidou, C.~Zhang, D.~Barducci, I.~Brivio, V.~Cirigliano and W.~Dekens, \textit{et al.}
[arXiv:1802.07237 [hep-ph]].

\bibitem{Ellis:2020unq}
J.~Ellis, M.~Madigan, K.~Mimasu, V.~Sanz and T.~You,
JHEP \textbf{04} (2021), 279
[arXiv:2012.02779 [hep-ph]].

\bibitem{Farina:2016rws}
M.~Farina, G.~Panico, D.~Pappadopulo, J.~T.~Ruderman, R.~Torre and A.~Wulzer,
Phys. Lett. B \textbf{772} (2017), 210-215
[arXiv:1609.08157 [hep-ph]].

\bibitem{Henning:2018kys}
B.~Henning, D.~Lombardo, M.~Riembau and F.~Riva,
Phys. Rev. Lett. \textbf{123} (2019) no.18, 181801
[arXiv:1812.09299 [hep-ph]].

\bibitem{Farina:2012xp}
M.~Farina, C.~Grojean, F.~Maltoni, E.~Salvioni and A.~Thamm,
JHEP \textbf{05} (2013), 022
[arXiv:1211.3736 [hep-ph]].

\bibitem{Dror:2015nkp}
J.~A.~Dror, M.~Farina, E.~Salvioni and J.~Serra,
JHEP \textbf{01} (2016), 071
[arXiv:1511.03674 [hep-ph]].

\bibitem{Azatov:2016sqh}
A.~Azatov, R.~Contino, C.~S.~Machado and F.~Riva,
Phys. Rev. D \textbf{95} (2017) no.6, 065014
[arXiv:1607.05236 [hep-ph]].

\bibitem{Degrande:2020evl}
C.~Degrande, G.~Durieux, F.~Maltoni, K.~Mimasu, E.~Vryonidou and C.~Zhang,
[arXiv:2008.11743 [hep-ph]].

\bibitem{Alwall:2014hca}
J.~Alwall, R.~Frederix, S.~Frixione, V.~Hirschi, F.~Maltoni, O.~Mattelaer, H.~S.~Shao, T.~Stelzer, P.~Torrielli and M.~Zaro,
JHEP \textbf{07} (2014), 079
[arXiv:1405.0301 [hep-ph]].

\bibitem{Degrande:2018fog}
C.~Degrande, F.~Maltoni, K.~Mimasu, E.~Vryonidou and C.~Zhang,
JHEP \textbf{10} (2018), 005
[arXiv:1804.07773 [hep-ph]].
\end{thebibliography}
\end{document}